# Versatile electronic properties and exotic edge states in single-layer tetragonal silicon carbides

Chao Yang[1,2], YueeXie[*1], Li-Min Liu[*2] and Yuanping Chen[1]

[1]Department of Physics,Xiangtan University,Xiangtan 411105,Hunan,China.

[2]Beijing Computational science Research Center,Beijing 100084,China

[*1]Corresponding author: xieyech@xtu.edu.cn

[*2]Corresponding author: limin.liu@csrc.ac.cn

## Abstract

Three single-layer tetragonal silicon carbides (SiC), termed as T1,T2 and T3, are proposed by density functional theory (DFT) computations. Although the three structures have the same topological geometry, they show versatile electronic properties from semiconductor (T1), semimetal (T2) to metal (T3).The versatile properties are originated from the rich bonds between Si and C atoms. The nanoribbons of the three SiCalso show interesting electronic properties. Especially, T1 nanoribbons possess exotic edge states, where electrons only distribute on one edge's silicon or carbon atoms. The band gaps of the T1 nanoribbons are constant because of no interaction between the edge states.

# Introduction

The discovery of graphene and its unique physical properties stimulated lots of interest to the fields of two-dimensional (2D) nanostructures.[1-4] Vast 2D nanosheets have been proposed based on various chemical elements or compounds, in which silicene,[5,6] $MoS_2$,[7,8] germanane,[9,10] etc. are synthesized successfully and new physical properties are found in them.[9,11,12] This further promotes the explorations of new 2D nanomaterials.

Although carbon and silicon both lie in the fourth column of the periodic table, their 2D nanosheets are different because of their completely different bonding characteristics. Silicene is a buckle structure while graphene is a perfect plane.[1,5] Therefore, it is interesting to study what will happen in the 2D nanosheet consisting of carbon and silicone compound. In the previous studies, some 2D silicon-carbon compounds including hexagonal rings have been proposed, such as hexagonal SiC,[13,14] g-$SiC_2$,[15] pt-SiC2,[16] $SiC_3$,[17] etc. These new structures show interesting properties such as large direct bandgap, improved photoluminescence, and high power conversion efficiency.

Recently, 2D structures consisting of rectangular rings have drawn much attention gradually. Liu et al.[18] proposed a T-graphene, a 2D tetra-symmetrical carbon sheet, and studied the electronic properties further. They found that the buckled T-graphene possesses a Dirac point while the planar structure is a metal. Li et al.[19] studied a 2D $MoS_2$ system with the repeated square-octagon rings and they found that the new $MoS_2$ allotrope possess both massless Dirac and heavy fermions near the Fermi surface. These studies indicate that the 2D tetra-symmetrical sheets have their own unique physical properties. To the best of our knowledge, 2D silicon carbides consisting of rectangular rings have never been studied up to date and thus it is nature to ask what new electronic properties do the 2D tetragonal SiC have.

In this paper, we proposethree 2DtetragonalSiCsheets,termed as T1, T2 and T3, as shown in Fig. 1. The cohesive energies and phonon dispersions of these 2Dsheets indicate that they are meta-stable SiCallotropes. Although the three structures have the same topological geometry, they exhibit completely different electronic properties,varying from metal to semimetal with Dirac cone and then to semiconductor. T1 consisting of Si-C bonds is a semiconductor. T2 and T3 consist of three kinds of bonds, i.e., Si-C, C-C and Si-Si. The former is a semimetal because of its two (Si and C) rectangular sublattices, while the later is a metal because of the C-C and Si-Si dimers between the rectangular rings. The electronic properties of nanoribbons of the 2D structures are also studied. We find that the band gaps of T1 nanoribbons are constant, and in these ribbons there exist exotic edge states whereelectrons only distribute on the (Si or C) edge atoms of one side.

## Models and Methods

The three 2DtetragonalSiCs, named as T1, T2 and T3, are shown in Figs. 1(a-c), respectively.All structures have the same topological geometry consisting of four-membered rings.The ratios of C and Si atoms are 1:1. Thefour-membered rings inT1 and T3have two C atoms and two Si atoms, while those in T2 are pure carbon rings or silicone rings.T1only consists of Si-C bonds while T2 and T3 consist of three kinds of bonds, C-C, Si-Si and Si-C. These tetragonal 2D SiCspossibly can be synthesized experimentally by using bottom-up method because of existing of C, Si and SiC four-membered moleculessuch as the $C_4H_4$,$Si_2C_2$, $Si_4$, $(Si_4H_4)^{2-}$, etc.[20-24]

The corresponding calculations were employing theVienna ab initio simulation package (VASP).[25]The interaction between ions and electrons is described by the projector-augmented wave (PAW) method as the plane-wave basis set with an energy cutoff of 450eV is used.[26]The PBE functional constructed by Perdew, Burke, and Ernzerhof are used to calculate the exchange-correlation energy.[27]The brillouin zone (BZ) is sampled with a $10\times10\times1$ Monkhorst-Pack special k-point grid including the Γ point. A vacuum region of 10 Å is added to the crystal plane to ensure that the wave functions vanish smoothly at the edge of the cell as an isolated system that would be required. The convergence criterion of the self-consistency process is set to $10^{-8}$eV between two ionic steps to calculate the phonon dispersion. The performed phonon calculations were using the Phonopy package[28] with the forces calculated by the VASP code.

## Results and Discussion

After geometryoptimized, the unit cells of the three single-layer SiCs in Fig. 1 maintain tetragonal form, but some differences among them are also exhibited. The detailedstructural parameters are given in Table 1. T1 and T2 are still planarstructures, while T3 becomes a buckled structure with the distance d≈0.48Å. The rectangular rings in T2 are still perfect, while those in T1 and T3 have a little distortion.The bond lengths of Si-C, C-C and Si-Si are also different in the three structures.

We first discuss the stability of the three 2D SiCs. The cohesive energies of T1, T2 and T3 are -6.77, -6.54 and -6.47 eV/atom, respectively, which are higher than that of the hexagonal 2D SiC (-7.03 eV/atom). On the other hand, we calculatethe phonon dispersions of the three structures, as shown in Figs. 2(a-c). One can find that there is no soft mode in the whole BZ, indicating the thermodynamic stability of the three structures. Therefore, T1, T2 and T3 are three new metastable SiC structures.

On the other hand, we calculatethe phonon dispersions of the three structures, as shown inFigs.2(a-c). One can find that there is no soft mode in the whole BZ, indicating the thermodynamic stability of the three structures. Therefore, T1, T2 and T3 are three new metastable SiC structures.

Figures 3(a-c) show the band structures and partial density of states (PDOS) of T1, T2 and T3, respectively.It is found that the three structures exhibit versatile electronic properties from semiconductor, semimetal to metal. T1 is a direct band-gap semiconductor with a gap 2.26eV [see Fig. 3(a)]. Boththe valence band maximum (VBM) and conduction bandminimum (CBM) are located between the K pathsΓ-X or Γ-M.ThePDOSin right panel of Fig.3(a) indicates that the band edge state around VBM is dominated by $p_z$orbital of C atoms,while that around CBM is dominated by $p_z$orbital of silicone atoms.To further exhibit the electronic properties of T1, we calculate the charge densities of the states at VBM and CBM as shown in Fig. 3(d). One can find that the band edge states are both induced by strong polar covalent bonds. The VBM is related to the polar covalentbond where electrons mainly distribute around the carbon atoms, while electrons at CBM mainly distribute around the silicon atoms. The existing of polar double bonds between the four-membered ringsis the reason why T1 is a semiconductor, like the case of single layer hexagonal SiC.[16, 29]

The band structure of T2in Fig.3(b) indicates that it is a semimetal like graphene, in which a Dirac point appears near the K point X.Our calculation indicates that the Fermi velocity around the Dirac point is about $10^5$ m/s close to that of graphene.The zero PDOS at the Fermi level further demonstrate the semimetallic property of T2. Similar to the case of T1, the bonds between the four-membered ringsare polar double bonds.Electrons around the Fermi level are localized on C atomsor Si atoms[see Fig. 3(e)]. However, T2 is a semimetal rather than semiconductor because the rectangular C ring and silicon ring form two sublattices.The polar double bonds are somewhat similar to the π and π* bonds in the graphene, and thusinduce two crossing bands at the Fermi level, i.e. forming the Dirac point.

T3 is a metal, as seen from Fig. 3(c) where several bands cross the Fermi level. The PDOS profile shows that the Fermi level and around are dominated by pzorbitals of carbon and silicon. We present the charge densities of two states at the Fermi level, as shown inFig. 3(f). One can find that the bonds in T3 are completely different from those in T1 and T2 because of T3's buckling structure. In the left panel of Fig. 3(f), there are σ bonds between silicon dimers and banana σ bonds between atoms silicon and carbon. A conducting network formed by these bonds leads to T3's metallic character. In the right panel of Fig. 3(f), big π bonds are observed between the carbon dimer and silicon atom, which also forms a conducting work. Additionally, it is noted that in many cases silicon or carbon dimers will result in metallic characterof SiC structures, such as single layer $SiC_2$ and reconstructed SiC surface.[16, 30-32]

Due to the versatile electronic properties of the three SiCs, it is interesting to explore the properties of their nanoribbons. The configurations fornanoribbonsofT1, T2 and T3 are shown in Fig. 4.Figures 4(a-c) represented the armchair nanoribbonsof T1, T2 and T3, while Figs. 4(d-f) represented their zigzag nanoribbons, respectively. The danglingbonds atthe edges are saturated byhydrogen atoms.The relation between the bandgap of these ribbons and their width is shown in Fig.5(a). The band gaps of T2 armchair nanoribbonsexhibit remarkableoscillating behaviorwiththe width increasing, while band gaps of T2 zigzag nanoribbons quickly converge to zero. Theoscillating behavior is similar to the case in the nanoribbons ofplanar $C_4$ carbon sheet.[18, 33] The nanoribbons with zero band gap are Dirac materials, indicating theyinherit the Dirac point from 2D sheet.For the T3 structure, nearly all nanoribbons are metal because of the metallic properties of 2D structure.This is similar to the pt-$SiC_2$ and planar T-graphenewhose 2D sheets and nanoribbonsare metallic.[16, 18]

Electronic properties of T1 nanoribbons are completely different from those of T2 and T3 nanoribbons [see Fig. 5(a)]. The band gaps of T1 nanoribbons are constant except small changes induced by quantum size effect at very narrow width.Theband gap of zigzag nanoribbonsis about 2.0eVwhile that of the armchair nanoribbonsis about 1.6eV. Figure5(b) displays the band structure of armchair nanoribbon with width $N_{1a}$=8. It is found that the valence and conduction bands closest the Fermi level are all flat bands. The charge densities in the insets of Fig. 5(b) illustrate that the electrons in the flat conduction bands are mainly localized on the carbon atoms of upper edge, while those in the flat valence bands are mainly localized on the silicon atoms of bottom edge. This demonstrates that the two edge states are related to the flat bands, which are different from the edge states in the graphene whose edge states are related to the Dirac cone.[34] For the widernanoribbons, the interactionsbetween the two edge states disappear, and thus the flat bands do not shift with the increase of width. The band gaps of nanoribbons are determined by the flat bands, so the band gap does not change with the width. The same case occurs in the zigzag nanoribbons (see Fig. 5(c)), however, the band gap of zigzag nanorbibbons is larger than that of armchair nanoribbons because of theanisotropyof T1.

## Conclusion

In summary, we proposed three single-layer tetragonalSiCs (T1, T2 and T3) and studied their electronic properties by using the density functional theory (DFT) computations.The analysis of stability indicates that the three structures are meta-stable allotropes of SiC. T1, T2 and T3 have the same topological geometry, but their electronic properties are very different. T1 is a direct band-gap semiconductor,T2is a semimetal whose band structure has a Dirac cone, while T3 is a metal.The versatile properties are originated from the rich bonds between Si and C atoms. The electronic properties of nanoribbons of the three 2D sheets are also studied. The results indicatedthe existence of exotic edge states related to the flat bands in the T1 nanoribbons.Electrons in these states only distribute on the silicon or carbon atoms of one edge. The band gaps of the T1 nanoribbons are constant because there is no coupling between the edge states in the wider nanoribbons.

# Acknowledgments

This work was supported by the National Natural Science Foundation of China (Nos. 51176 161, 51376005 and, 11474243).

Table 1 The structural parameters for T1, T2, T3 and hexagonal 2D SiC, respectively.

Figure 1 Atomic structures for (a) T1, (b) T2, (c) T3, top and bottom panelsare top- and side-view structures. Blue and black ballsrepresent Siand C atoms, respectively.T1 and T2 are planarstructures, while T3 is a buckled structure with the distance d ≈0.48Å. The angle parameters areα: 85.6°, β: 94.4°, γ: 92.0°, δ: 88.0°.

Figure 2Phonon dispersions of (a) T1, (b) T2 and (c) T3, respectively.

Figure 3Band structure and PDOS(right panel)for (a) T1, (b) T2 and (c) T3. (d) Partial charge densities for the states at VBM (left panel) and CBM (right panel). (e) Partial charge densities for the two states at the Dirac point. (f) Partial charge densities for the two states at the Fermi level. Left panel corresponds to the point 1 in (c), while right panel corresponds to the point 2 in (c),respectively.

Figure 4 Armchair nanoribbons for (a) T1, (b) T2 and (c) T3. (d), (e) and (f) are zigzag nanoribbons for T1, T2 and T3, respectively. Blue and black balls represent Si and C atoms, while the red ballsrepresent the hydrogen atoms. $N_{1a/1z}$, $N_{2a/2z}$and $N_{3a/3z}$ are widths of T1, T2 and T3 armchair/zigzag nanoribbons, respectively.

Figure 5 (a) Bandgaps of T1,T2 and T3 nanoribbonsas a function of their widths. (b) Band structure of T1armchairnanoribbon with $N_{1a}=8$.(c)Band structure of T1zigzagnanoribbon with $N_{1z}=8$.Inset: Partial charge densities for the states in the flat bands.

| structure | C-C length | Si-C length | Si-Si length | angle | planar or buckling |
|---|---|---|---|---|---|
| hexgonal | | 1.79 Å | | | planar |
| T1 | | $d_1$: 1.81 Å | | α: 94.4° | planar |
| | | $d_2$: 1.75 Å | | β: 85.6° | |
| T2 | 1.46 Å | 1.78 Å | 2.26 Å | | planar |
| T3 | 1.37 Å | 1.84 Å | 2.24 Å | γ: 92.0° | buckling |
| | | | | δ: 88.0° | |
| $d_1$ ($d_2$): length of Si-C bond in (between) the four-membered ring. | | | | | |

Table 1

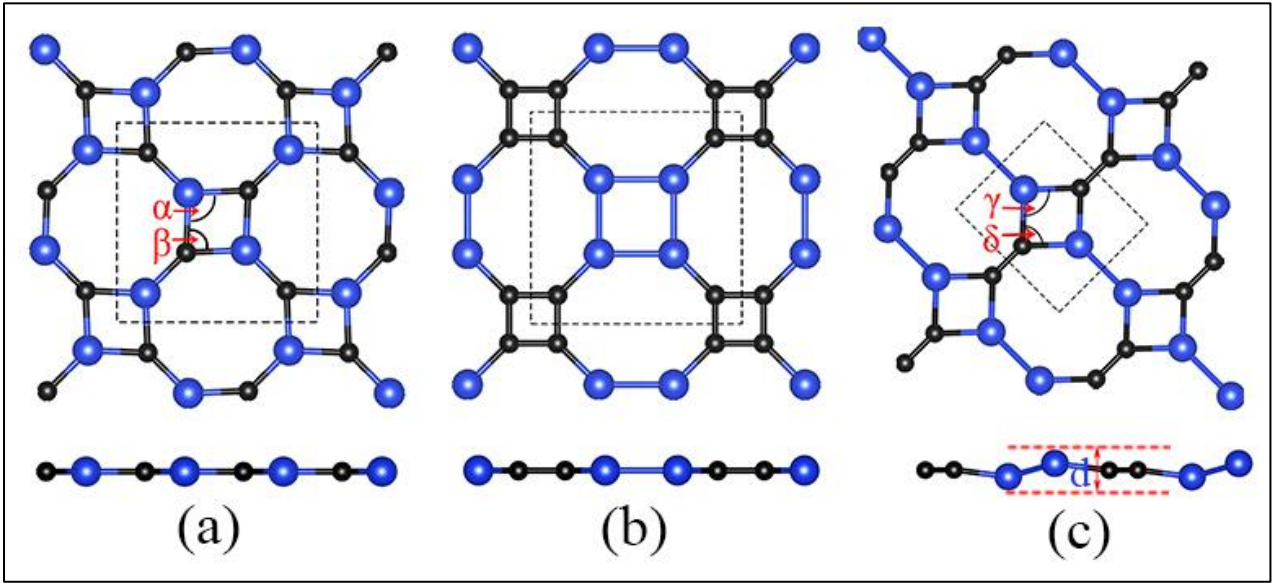

Figure 1

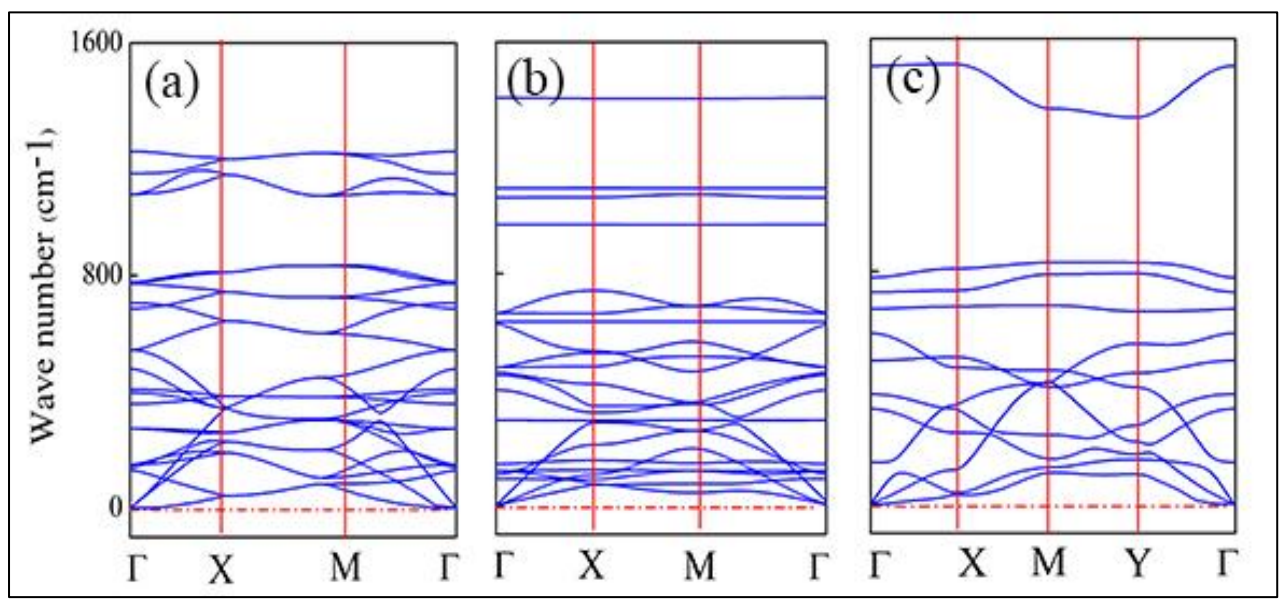

Figure 2

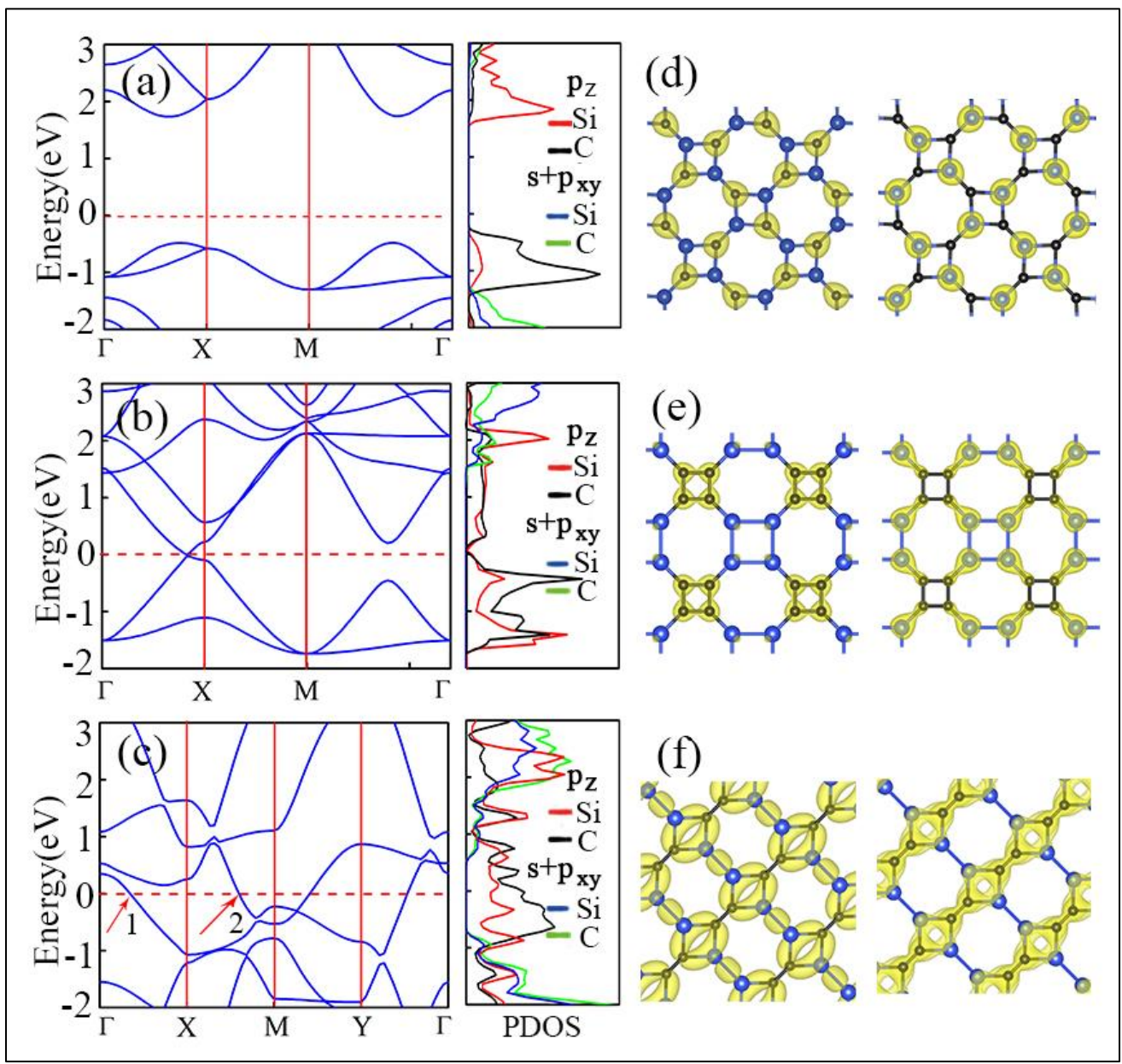

Figure 3

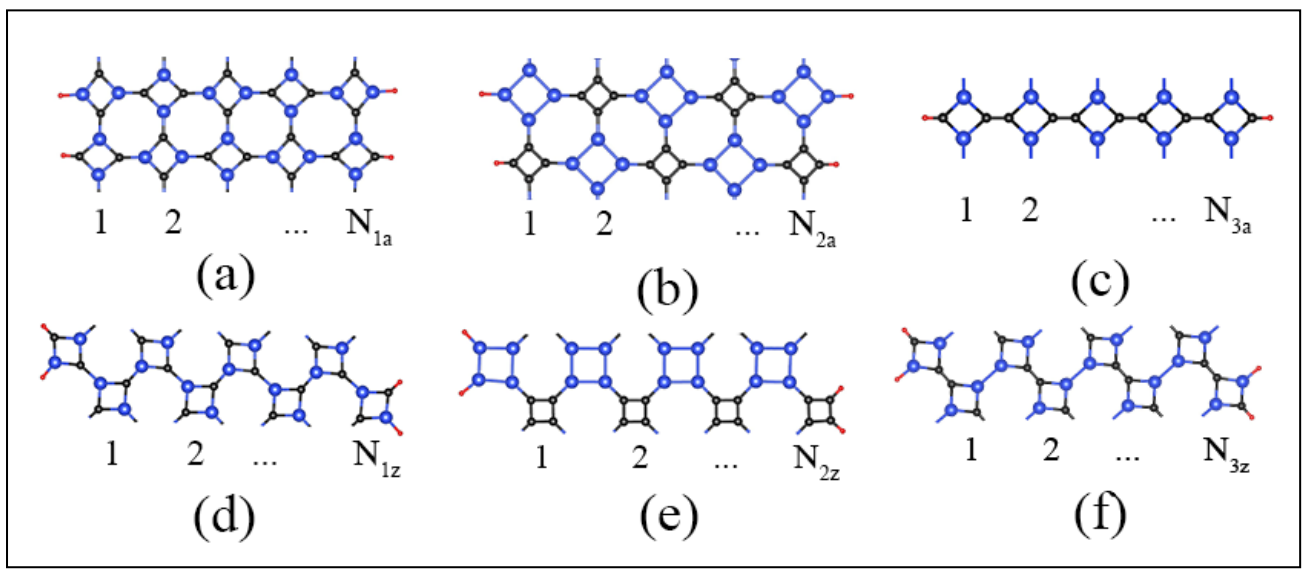

Figure 4

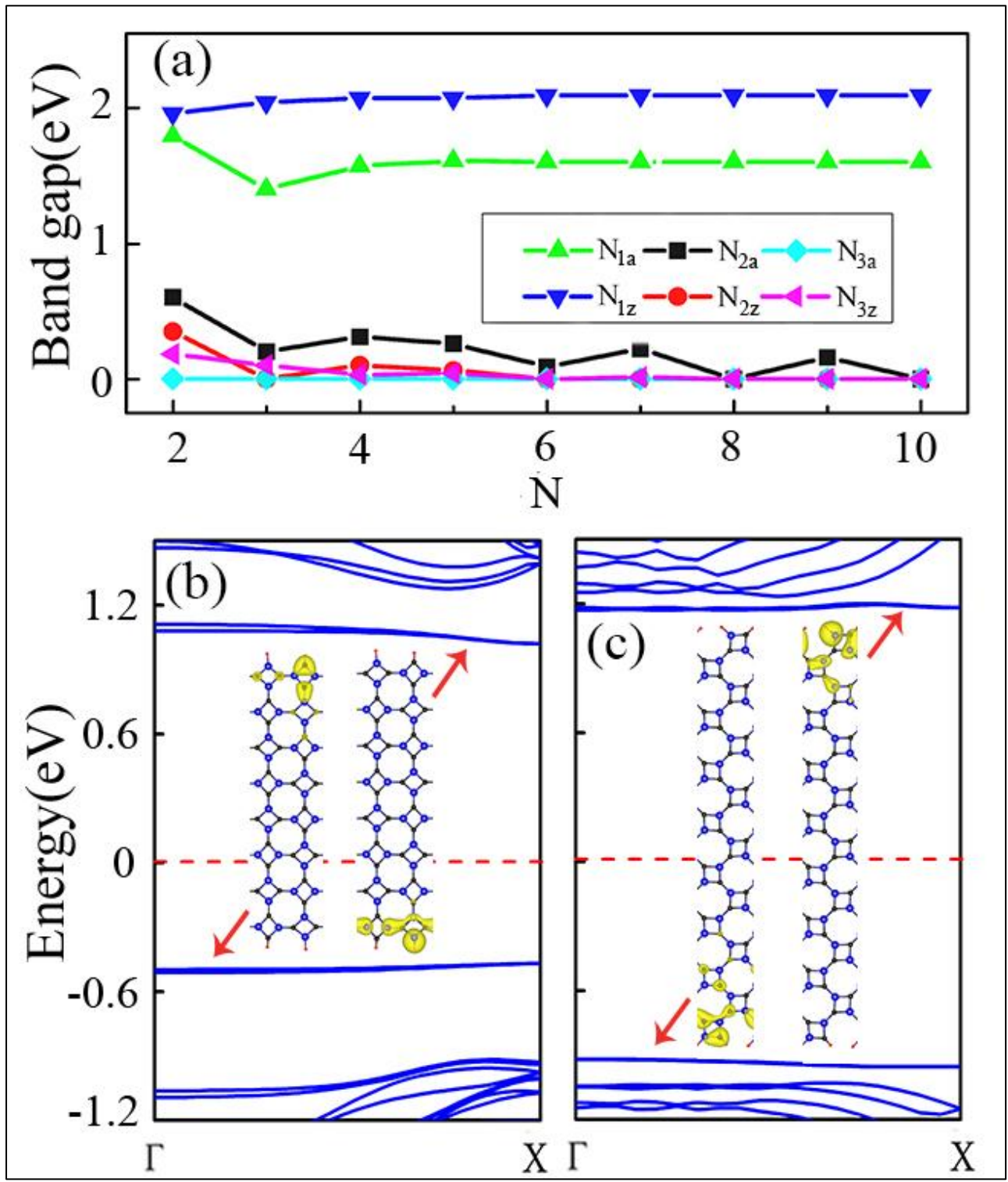

Figure 5